\documentclass[preprint,3p,times, twocolumn]{elsarticle}

\usepackage{graphicx}

\journal{Physics Letters B}

\begin{document}

\begin{frontmatter}

\title{Precise determination of quadrupole and hexadecapole deformation parameters of the $sd$-shell nucleus, $^{28}$Si}

\author[BARC,HBNI]{Y. K. ~Gupta\corref{cor1}}
\cortext[cor1]{Corresponding author, Email:ykgupta@barc.gov.in}

\author[SVNIT]{V. B. Katariya}
\author[BARC]{G. K. Prajapati}
\author[KYOTO]{K. Hagino}
\author[SVNIT]{D. Patel}
\author[IITR]{V. Ranga}
\author[BARC,HBNI]{L. S. Danu}
\author[BARC]{A. Pal}
\author[BARC]{B. N. Joshi}
\author[TIFR]{S. Dubey}
\author[TIFR]{V. V. Desai}
\author[IITR]{S. Panwar}
\author[ND]{U. Garg}
\author[BARC]{N. Kumar}
\author[BARC]{S. Mukhopadhyay}
\author[BARC,HBNI]{Pawan Singh}
\author[BARC]{N. Sirswal}
\author[PU]{R. Sariyal}
\author[TIFR]{I. Mazumdar}
\author[BARC]{B. V. John}

\address[BARC]{Nuclear Physics Division, Bhabha Atomic Research Centre, Mumbai - 400085, India}
\address[HBNI]{Homi Bhabha National Institute, Anushaktinagar, Mumbai 400094, India}
\address[SVNIT]{Department of physics, Sardar Vallabhbhai National Institute of Technology, Surat -395007, India}
\address[KYOTO]{Department of Physics, Kyoto University, Kyoto 606-8502, Japan}
\address[IITR]{Department of Physics, IIT Roorkee, Roorkee-247667, India}
\address[TIFR]{Tata Institute of Fundamental Research, Mumbai 400005, India}

\address[ND]{Department of Physics and Astronomy, University of Notre Dame, Notre Dame, IN 46556, USA}
\address[PU]{Department of Physics, Panjab University, Chandigarh-160014, India}


\date{\today}

\begin{abstract}
Quasi-elastic (QEL) scattering measurements have been performed using $^{28}$Si projectile off a $^{90}$Zr target at energies around the Coulomb barrier.  A Bayesian analysis within the framework of coupled channels (CC) calculations is performed in a large parameter space of quadrupole and  hexadecapole deformations ($\beta_{2}$ and $\beta_{4}$) of $^{28}$Si. Our results clearly show that $^{28}$Si is an oblate shaped nucleus with  $\beta_{2}$=-$0.38 \pm 0.01$ which is in excellent agreement with electromagnetic probes. A precise value of hexadecapole deformation for $^{28}$Si, $\beta_{4}$=+$0.03 \pm 0.01$, along with a consistent value of quadrupole deformation has now been determined for the first time using QEL scattering. A remarkable agreement between the experimental  $\beta_{4}$ value of $^{28}$Si and  Skyrme-Hartree-Fock based calculations is obtained. The QEL results obtained previously for $^{24}$Mg (prolate) and the present result for $^{28}$Si (oblate) hereby affirm the strong sensitivity of the quasi-elastic scattering to ground state deformations, thus reinforcing its suitability as a potential probe for rare exotic nuclei.
\end{abstract}

\end{frontmatter}

Nuclear deformation presents a fascinating example of a delicate balance between liquid drop nature and quantum many body aspects of the strongly interacting nucleons \cite{bohr1969nuclear}. In the vicinity of the $\beta$-stability line, the nuclei close to the major shell closures exhibit spherical ground state shapes. While moving away from the shell closures, the shell structure of valence nucleons primarily governs the ground state shapes \cite{casten1990nuclear, Mackintosh1977, Bender2003}. With the advent of RIB factories along with the associated development of more sophisticated experimental techniques and parallel advancement in modern theoretical frameworks, a significant interest in studying the shell structure evolution in the regions of extreme isospin, far from the valley of $\beta$-stability, is among the primary foci of current nuclear physics research \cite{Otsuka2020}.

Dominantly, the shell structure of valence nucleons leads to the axially symmetric deformations with reflection symmetry, namely the quadrupole and the hexadecapole \cite{Mackintosh1977}. The quadrupole deformation ($\beta_{2}$) is most commonly experienced as either elongated (prolate) or prostrated (oblate) shape. The sign and value of $\beta_{2}$ is now being determined for many of the unstable nuclei \cite{Glasmacher1998}, but, the higher order hexadecapole deformation ($\beta_{4}$) is not precisely known even for a majority of the  stable deformed nuclei \cite{ykgplb2020, StephensPRL1971, BRUCKNER1974159, GKaur2018}.

An island of deformed nuclei is known to exist in the $sd$ shell stable region for the past several decades \cite{IbbotsonPRL1998}. In this mass region, the sign of quadrupole deformation (prolate versus oblate) is ambiguous for many of the nuclei and the knowledge of the hexadecapole deformation is practically non-existent. In recent times, in connection to $N$=28 shell quenching \cite{42SiPRL2019, suzukiPRC2021}, a variety of calculations have been performed for nuclear deformation and its causes in the wide isotopic chains of Mg, Si, S, and Ar ($sd$ and $fp$ shells) employing different formalisms and density functionals \cite{DaoPRC2022, WernerNPA1996, BenderPRC2008}.  These calculations predict  a broad potential energy curve as a function of $\beta_{2}$ for the $sd$ shell nuclei including the stables ones \cite{GaamouciPRC2021, WinHaginoPRC2008, LiPTEP2013, Dhiman2021, MeiHaginoPRC2018}. It is crucial to learn about the ground state shapes in the stable region at first to rely on the predictive power of modern theories for exotic nuclei being investigated at RIB factories, where beam intensity is a serious concern.

In case of the $sd$ shell nucleus, $^{28}$Si, the aforementioned theories predict conflicting quadrupole shapes, varying from oblate to prolate, with generally no predictions about the hexadecapole deformation \cite{WinHaginoPRC2008, LiPTEP2013, WernerNPA1996}. Earlier, inelastic scattering probes such as the  electron-scattering \cite{Cooper1976,Horikawa1971}, Coulomb excitation \cite{FEWELL1979, BALLNPA1980, Haouat1984}, proton-scattering \cite{Cole1966, Blairppprime1970, swiniarski1969, Leo1979}, neutron-scattering \cite{Leo_neutron_28Si, Bottcher1983, Haouat1984}, deuteron-scattering \cite{Niewodniczanski1964}, $\alpha $-scattering \cite{KokameAlpha, Rebel1972, Harakeh1979}, and heavy-ions \cite{Dudek1978, Mittig1974} have been employed to estimate the $\beta_{2}$ and $\beta_{4}$ in the sd-shell region.  Majority of the measurements are in favor of an oblate ground state shape of $^{28}$Si, however, with a large variation in its $\beta_{2}$ value from -0.34 to -0.55 \cite{BALLNPA1980, Horikawa1971, swiniarski1969, Blairppprime1970, Haouat1984, Bottcher1983, Dudek1978}. In fact, some of the experimental investigations show it to be a prolate shaped nucleus \cite{Cole1966, Leo1979, Niewodniczanski1964, KokameAlpha}. Its $\beta_{4}$ value determined so far varies over a quite large range.


\begin{figure}[t]\includegraphics[trim= 0.2mm 0.2mm 0.2mm -2mm, clip, height=0.5\textheight]{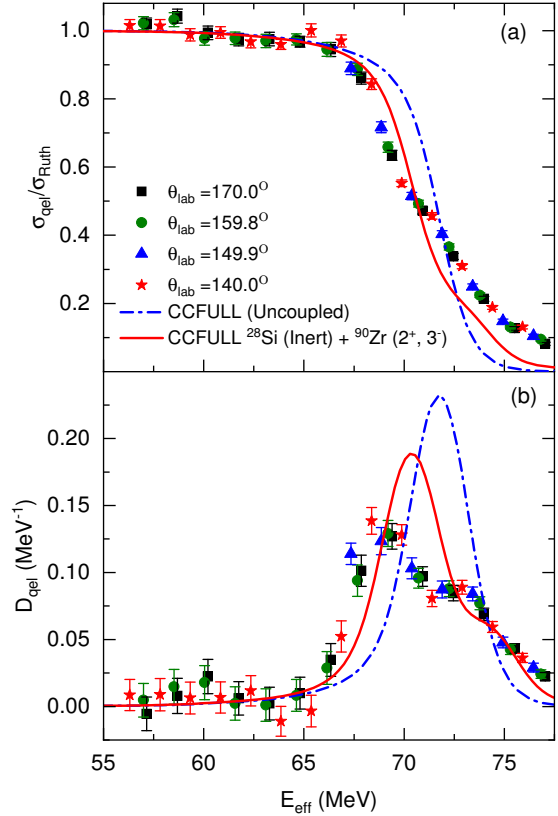}
\caption{\label{28Si_Expt_BD} Quasi-elastic excitation function (panel (a)) and extracted  barrier distribution (panel (b)) at four backward angles for $^{28}$Si + $^{90}$Zr reaction. The dash-dotted and  solid lines represent CCFULL calculations without including any coupling (uncoupled) and with including vibrational couplings  of $^{90}$Zr (2$^{+}$, 3$^{-}$), respectively.}
\end{figure}

During the heavy-ion fusion, the  coupling of internal degrees of freedom of the fusing nuclei, such as vibrational (spherical), rotational (deformed), and particle transfer, gives an opportunity to gain insight about the nuclear structure \cite{Dasgupta, HaginoPTP2012} from the measured Fusion Barrier Distributions (FBDs). Following the above idea, recently a very precise value of hexadecapole deformation was determined for a prolate shaped, another $sd$ shell nucleus, $^{24}$Mg using Quasi-elastic (QEL) scattering  measurements \cite{ykgplb2020}. The QEL scattering provides an alternate route to get a good representation of FBD with some additional advantages \cite{Piasecki2005}. In the present letter, results obtained on ground state $\beta_{2}$ and $\beta_{4}$ values for $^{28}$Si nucleus using QEL scattering off $^{90}$Zr target, are presented. The $\beta_{2}$ value determined in the present work rule-out many predictions of modern theories which show a broad potential energy curve as a function of $\beta_{2}$. The ground state hexadecapole deformation along with a consistent value of quadrupole deformation of $^{28}$Si has been determined precisely for the first time. Present results for $^{28}$Si and those reported earlier for $^{24}$Mg affirm the strong sensitivity of the quasi-elastic scattering probe with respect to higher order hexadecapole deformation, that could be of significant use for exotic nuclei using poor intensity RIBs.

Quasi-elastic measurements were carried out using $^{28}$Si DC beam from BARC-TIFR 14 MV Pelletron accelerator facility. Highly enriched ($>$95\%) $^{90}$Zr (150 $\mu$g/cm$^{2}$) deposited in oxide form on $^{12}$C (40 $\mu$g/cm$^{2}$) was used as the target. Quasi-elastic events were detected using four very thin ($\simeq$ 15 $\mu m$) silicon surface barrier (SSB) detectors placed at 140.0$^{\circ}$, 149.9$^{\circ}$, 159.8$^{\circ}$, and 170.0$^{\circ}$ with respect to the beam direction. The angular opening of each detector was restricted to close to $\pm 1^{\circ }$. Two more SSB detectors (1 mm), were mounted at 20.0$^{\circ}$ in the reaction plane on either side of the beam direction for the purpose of Rutherford normalization. Each of these monitor detectors had a collimator of 1 mm diameter.  Rutherford scattering peak of $^{90}$Zr was well separated from that of the  $^{12}$C (backing) and $^{16}$O (target is ZrO$_{2}$) at forward angles ($\pm$20$^{\circ}$).

Beam energies were used in the range of 70 to 102 MeV in steps of 2-MeV. At every change of beam-energy, the transmission of the beam was maximized through a collimator of 5 mm diameter, enabling a hallo-free beam. The solid-angle ratios of monitor to back-angle thin detectors were experimentally determined from  Rutherford scattering of $^{28}$Si projectile off $^{197}$Au (150 $\mu$g/cm$^{2}$) target at 70 and 72 MeV beam energies.  The quasi-elastic events consist of elastic, inelastic from projectile and/or target excitations and to some extent from transfer events. Most of the QEL events stopped within the thin SSB detectors placed at the backward angles. Owing to high negative $Q$-values in $^{28}$Si+$^{90}$Zr reaction, the transfer channels are not expected to contribute in QEL events. In any case, all the QEL events consist of Projectile Like Fragments (PLFs), stopped within the thin SSBs,  were clearly separated from evaporated Light Charged Particles (LCPs) from pulse height analysis. The two-body kinematics did now allow to have any contribution from $^{12}$C and $^{16}$O, present in the target. Among quasi-elastic events, the elastic events were dominant. All the SSB detectors were energy calibrated using elastic peaks of different beam energies. Successive changes in the kinetic energies of elastic events with varying beam energy were in agreement with two-body kinematics at all the angles in going from from 140$^{\circ}$ to 170$^{\circ}$, which further benchmarked the identification of quasi-elastic events. The beam energies were corrected for energy loss in the half-thickness of the target.

Differential cross section for quasi-elastic events at each beam energy was normalized with Rutherford scattering cross section. The center-of-mass energy ($E_\mathrm{c.m.}$) was corrected for centrifugal effects at each angle as follows \cite{Piasecki2005, Timmers1995, BKN2007, }:
\begin{equation}
E_\mathrm{eff} =\frac{2E_\mathrm{c.m.}}{(1+\mathrm{cosec}(\theta_\mathrm{c.m.}/2))}
\end{equation}

where $\theta_\mathrm{c.m.}$ is the center-of-mass angle.  The quasi-elastic excitation function for the $^{28}$Si + $^{90}$Zr reaction is shown in the Fig. \ref{28Si_Expt_BD}(a) at the four backward angles. It is seen that the quasi-elastic excitation functions at different laboratory angles join quite smoothly. Quasi-elastic barrier distribution $D_\mathrm{qel}$ ($E_\mathrm{eff}$) from quasi-elastic excitation function was determined using the relation \cite{Timmers1995}:
\begin{equation}
D_{qel}(E_\mathrm{eff}) = -\frac{d}{dE_\mathrm{eff}}\bigg[ \frac{d\sigma_\mathrm{qel}(E_\mathrm{eff})}{d\sigma_\mathrm{R}(E_\mathrm{eff})}\bigg],
\end{equation}
where $d\sigma_\mathrm{qel}$ and $d\sigma_\mathrm{R}$ are the differential cross sections for the quasi-elastic and Rutherford scatterings, respectively. A point difference formula is used to 
evaluate the barrier distribution, with the energy step of 2 MeV in the laboratory frame of reference. Similar to the excitation function, the barrier distribution determined from excitation functions at different laboratory angles joins quite smoothly as shown in the Fig. \ref{28Si_Expt_BD}(b). The smooth joining of the data in excitation function as well as derived barrier distribution, ensures correct identification of the quasi-elastic events.

\begin{figure}\includegraphics[trim= 0.2mm 0.2mm 0.2mm -2mm, clip, height=0.30\textheight]{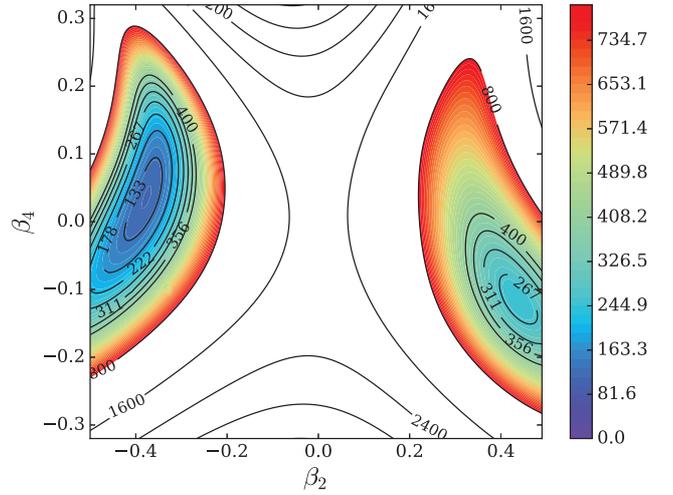}
\caption{\label{Chisq} A $\chi^{2}$ distribution in the two dimensional space of $\beta_{4}$ and $\beta_{2}$ of $^{28}$Si, determined by comparing experimental QEL excitation function with CC calculations (see text).}
\end{figure}


Coupled channels (CC) calculations were carried using a modified version of CCFULL code \cite{ccqel} for quasi-elastic scattering (see Ref. \cite{ykgplb2020} for details). The CC calculations were carried out at  first without including any channel coupling. These uncoupled CC calculations are represented by the dash-dotted lines in Figs. \ref{28Si_Expt_BD}(a) and (b). It is clearly seen that uncoupled calculations cannot reproduce the experimental data. The CC calculations were further performed  including the vibrational couplings of the  target, $^{90}$Zr, while the projectile nucleus, $^{28}$Si was assumed to be inert. Vibrational quadrupole (2$^{+}$) state at 2.18 MeV and the octupole (3$^{-}$) state at 2.75 MeV of $^{90}$Zr, were taken into account. The coupling strengths for the above 2$^{+}$ and 3$^{-}$ states of $^{90}$Zr were used as 0.089 and 0.211, respectively, as determined earlier \cite{ykgplb2020}. It is clearly seen that the CC calculations only with the vibrational couplings of $^{90}$Zr deviates significantly from experimental data, raising the urge to include other degrees of freedom of either projectile or the target within the CCFULL framework.

$^{28}$Si shows a rotational band built on the ground state \cite{NNDC} with a non-zero quadrupole moment, $Q(2^{+})$ \cite{Stone2016}. These features suggest an importance for including the rotational couplings of $^{28}$Si within the CC calculations in order to reproduce the quasi-elastic excitation function and the barrier distribution for $^{28}$Si + $^{90}$Zr reaction. Rigid rotor model was used for this purpose. Along with quadrupole deformation, the hexadecapole  deformation in the ground state of $^{28}$Si was also considered to be included in the CC calculations. In order to reproduce the experimental data and determine precise values of $\beta_{2}$ and $\beta_{4}$, the CC calculations were carried out in a large parameter space of quadrupole and hexadecapole deformations. The parameters $\beta_{2}$ and $\beta_{4}$ were varied in the range of -0.5 (oblate) to +0.5 (prolate) and -0.4 to +0.4, respectively with a step size of 0.01 for both the parameters. Coulomb and nuclear parts for both quadrupole and hexadecapole deformations were kept at same values. First three rotational states of $^{28}$Si, namely,  0$^{+}$, 2$^{+}$, and 4$^{+}$, were included in the CCFULL calculations. The coupling to the 6$^{+}$ state has been confirmed to give a negligible contribution. Large number of CC calculations were performed using the ``ANUPAM" supercomputer of BARC.
\begin{figure}\includegraphics[trim= 0.2mm 0.2mm 0.2mm -2mm, clip, height=0.27\textheight]{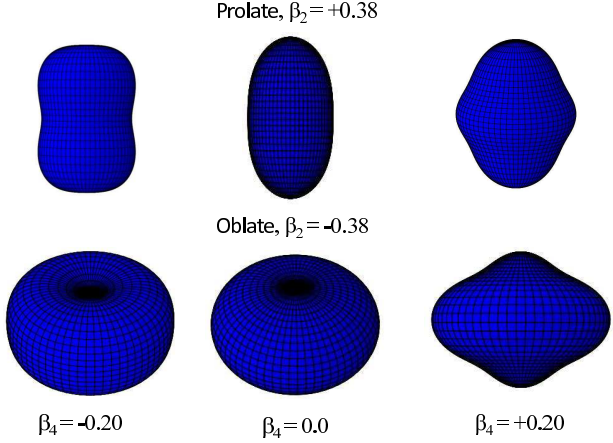}
\caption{\label{Shapes} Schematic shapes: top row for prolate ($\beta_{2}$=+0.38) and bottom row for oblate ($\beta_{2}$=-0.38) quadrupole deformation with different values of hexadecapole deformation ($\beta_{4}$).}
\end{figure}

\begin{figure}
\includegraphics[trim= 0.2mm 0.2mm 0.2mm -2mm, clip, height=0.35\textheight, angle=0]{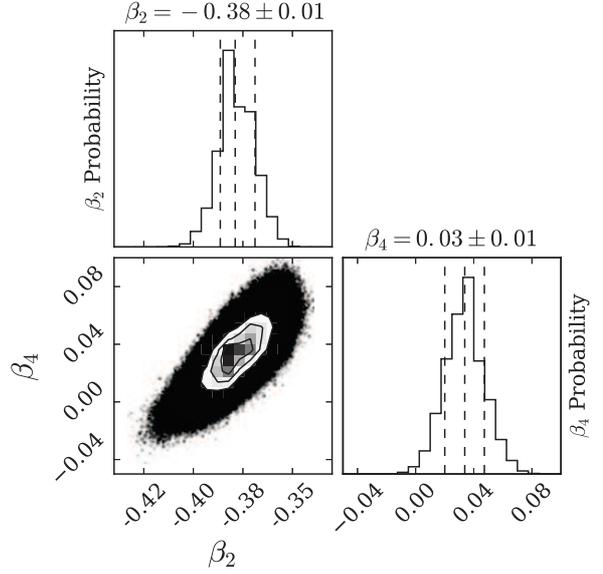}
\caption{Two-dimensional probability distributions for $\beta_2$ and $\beta_4$ of for $^{28}$Si, resulting from the MCMC simulation \cite{ykgplb2020} from the experimental data (see text). Plus- and minus-uncertainties are shown.}
\label{probability_dists}
\end{figure}

A $\chi^{2}$ was calculated between the experimental QEL scattering excitation function and CC calculations for each combination of $\beta_{2}$ and $\beta_{4}$ using following equation;
\begin{equation}
\chi^{2}(\beta_{2},\beta_{4})=\sum_{i=1}^{N}\frac{[Y_{i}-f(\beta_{2},\beta_{4})]^2}{\sigma_{i}^{2}}
\label{Chi}
\end{equation}

where $Y_{i}$ represents the experimental value of the excitation function at the $i^{th}$ energy point, $\sigma_{i}$ is the uncertainty in the data, and $f(\beta_{2},\beta_{4})$ represents corresponding CCFULL calculation for a particular combination of $\beta_{2}$ and $\beta_{4}$. In  Eq. \ref{Chi}, the summation runs over all the data points ($N$) in the effective energy $E_\mathrm{eff}$  range of 64 to 75 MeV. The $\chi^{2}$-distribution thus obtained in the two-dimensional space of $\beta_{4}$ and $\beta_{2}$ is shown in Fig. \ref{Chisq}. It shows two minimas,  one corresponding to oblate (left contour) and the other for prolate shape (right contour). The $\chi^{2}$ value corresponding to the oblate shape is approximately four times smaller than that for prolate shape as can be seen from the Fig. \ref{Chisq}. Thus, the $\chi^{2}$ distribution reveals an oblate ground state shape of $^{28}$Si with a certain non-zero value of the hexadecapole deformation. In order to visualize oblate and prolate shapes with positive and negative hexadecapole deformations, schematic shapes were generated as shown in Fig. \ref{Shapes} using following expression \cite{StephensPRL1971} for nuclear radius in a ``PYTHON" script;
\begin{equation}
R(\theta, \phi)= R_{o}\left[1+ \beta_{2}Y^{0}_{2}(\theta, \phi)+ \beta_{4}Y^{0}_{4}(\theta, \phi)\right],
\label{Radius}
\end{equation}
where, $R_{o}$=1.2$A^{1/3}$ fm, and $Y^{0}_{2}$ and $Y^{0}_{4}$ are the spherical harmonics for $L$=2 and 4, respectively. One can notice dramatic change in the shape of a nucleus with changing $\beta_{2}$ and $\beta_{4}$.

\begin{table}
\caption{\label{table1}Quadrupole and hexadecapole deformation of $^{28}$Si using different experimental probes. Results of theoretical calculations based on Skyrme-Hartree-Fock (SHF)  method \cite{SkyAx} are also shown.}
\begin{tabular} {lll}
\hline

\hline
\\
\bf{Experiment} &&\\
Present work                                     & -0.38$\pm$0.01             & +0.03$\pm$0.01\\
(CE\footnotemark[1])      \cite{BALLNPA1980}                     & -0.39                      &    \\
(e, e$'$ )\cite{Horikawa1971}                    & -0.39                      & +0.10\\
(n, n$'$ )\cite{Bottcher1983}                    & -0.39                      &   \\
(n, n$'$ )\cite{Haouat1984}                      & -0.48                      &  0.15 \\
(n, n$'$ )\cite{Haouat1984}                      & -0.42 $\pm$0.02            & +0.20 $\pm$ 0.05 \\
(n, n$'$ )\cite{Leo_neutron_28Si}                & +0.41                      &   \\
(p, p$'$ )\cite{swiniarski1969}                  & -0.34                      & +0.25 \\
(p, p$'$ ) \cite{Blairppprime1970}               & -0.55                      & +0.33 \\
(p, p$'$ ) \cite{Cole1966}                       & +0.41                      &  \\

(d, d$'$ )\cite{Niewodniczanski1964}             & +0.45                      &   \\
($\alpha$, $\alpha '$ )\cite{KokameAlpha}        & +0.36                     & \\
($\alpha$, $\alpha '$ )\cite{Rebel1972}               & -0.32 $\pm$ .01           &  +0.08$\pm$0.01\\
($^{16}$O, $^{16}$O$'$ )\cite{Dudek1978}         & -0.34                     &   \\

\\
\bf{Theory} &&\\

SHF-SV-min\cite{SV-min}  &-0.327 & +0.035\\
SHF-SV-bas\cite{SV-bas}  &-0.334 & +0.041\\
SHF-SLy4  \cite{SLy4}    &-0.333 & +0.047\\

\hline

\end{tabular}
\footnotemark[1]{Coulomb Excitation} 
\end{table}

\begin{figure}[t]\includegraphics[trim= 0.2mm 0.2mm 0.2mm -2mm, clip, height=0.45\textheight]{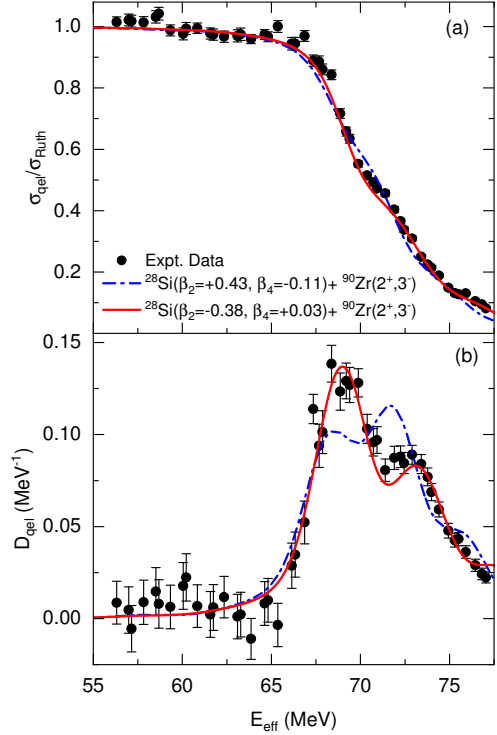}
\caption{\label{BD_FinalB2B4}  Quasi-elastic excitation function (panel (a)) and derived barrier distribution (panel (b)) for $^{28}$Si + $^{90}$Zr reaction. The dash-dotted (blue) and solid lines (red) show CC calculations for the two parameter sets, corresponding to prolate and oblate shapes, respectively. Vibrational couplings of $^{90}$Zr (2$^{+}$, 3$^{-}$) were included (see text).  }
\end{figure}

In order to obtain the quantitative values of $\beta_{2}$ and $\beta_{4}$ for $^{28}$Si and their associated uncertainties from present data of quasi-elastic excitation function, a Bayesian analysis with a Markov-Chain Monte Carlo (MCMC) framework was carried out as discussed in details in our previous work \cite{ykgplb2020}. Probability distributions obtained from the Bayesian analysis corresponding to the global minima (oblate) in the $\chi^{2}$-distribution are shown in Fig. \ref{probability_dists}. The $\beta_2$ and $\beta_4$ are moderately correlated with a correlation of $\sim +0.138$, which is evidenced graphically within the two-dimensional probability distribution as shown in Fig. \ref{probability_dists}. Examination of the projections of the probability density onto the parameter axes yields extracted values of $\beta_2 = -0.38 \pm 0.01$ and $\beta_4 = +0.03 \pm 0.01$, with approximately symmetric distributions centered at the medians. 

Experimental data for quasi-elastic excitation function and derived barrier distribution are compared with CC calculations using  the $\beta_{2}$ and $\beta_{4}$ parameters of $^{28}$Si, determined using the Bayesian analysis, as shown in Fig. \ref{BD_FinalB2B4}.  Data are also compared with CC calculations using $\beta_{2}$=+0.43 and $\beta_{4}$=-0.11, corresponding to the prolate shape minima in the $\chi^{2}$-distribution (see Fig. \ref{Chisq}). The quasi-elastic excitation function (Fig. \ref{BD_FinalB2B4} (a)) does not show a huge difference, but the derived barrier distribution which provides a finger print of the fusing partners, exhibits an enlarged sensitivity with the two parameter sets as depicted in the Fig. \ref{BD_FinalB2B4}(b). One can see that the agreement between CC calculations with the oblate shape of $^{28}$Si and the experimental data is excellent, whereas the calculations corresponding to the prolate shape deviate from the data quite significantly. Thus, Bayesian analysis carried out in the present work for QEL scattering data uniquely shows that $^{28}$Si is an oblate shaped in its ground state having $\beta_2 = -0.38 \pm 0.01$. Moreover, it also yields a precise value of hexadecapole deformation, $\beta_4 = +0.03 \pm 0.01$. 


The $\beta_{2}$ and $\beta_{4}$ values of $^{28}$Si have been reported in literature from the measurements of inelastic scattering off electron-, proton, neutron, deuteron, $\alpha$, $^{16}$O, and Coulomb excitation (CE). These values are listed in the Table  \ref{table1}.  One can see that previously determined ground state $\beta_{2}$ values of $^{28}$Si vary in a quite wide range. 
However, the electromagnetic probes, Coulomb excitation and electron scattering have been the primary tools to estimate the nuclear size and shapes. The $\beta_2$ value of $^{28}$Si determined in present work shows excellent agreement with those determined using electromagnetic probes (see Table \ref{table1}). Similarly, the $\beta_{2}$ value for $^{24}$Mg,  determined previously using quasi-elastic scattering shows excellent agreement with that determined using electron scattering (see Ref. \cite{ykgplb2020}). Apparently, Coulomb excitation and electron scattering do not show a good sensitivity 
to hexadecapole deformation, neither for $^{24}$Mg nor for $^{28}$Si. The $\beta_4$ of $^{28}$Si determined earlier vary quite significantly from +0.08 to +0.33. It is the first time that $\beta_4$ of $^{28}$Si has been determined precisely to be +0.03$\pm$0.01 along with precise $\beta_2$ value having good overlap with those determined using electromagnetic probes (see Table \ref{table1}). 

Table \ref{table1} also shows the results of Skyrme-Hartree-Fock calculations obtained with the computer code SkyAx \cite{SkyAx} for three different parameter sets, SLy4 \cite{SLy4}, 
SV-min \cite{SV-min}, and SV-bas \cite{SV-bas}. The deformation parameters $\beta_2$ and $\beta_4$ are evaluated with the calculated Q$_{2}$ and Q$_{4}$ moments by taking into account the non-linear 
terms of the deformation parameters. Even though the absolute values of $\beta_2$ are somewhat smaller than the value obtained in this analysis, it is remarkable that 
the sign and the value of $\beta_4$ are consistent with the present results.


In the rare earth region, where coupling effects are much stronger than the lighter mass, the quasi-elastic scattering measurements have been used to determine the hexadecapole deformation for some of the nuclei \cite{Jia2014}.  Along with our recent results for $^{24}$Mg (prolate), the present results for  $^{28}$Si (oblate) demonstrate that quasi-elastic scattering is a potential probe to determine the ground state deformation parameters in the lighter mass region also such as the $sd$- and $fp$ shell regions.

In summary, quasi-elastic measurements have been carried out in $^{28}$Si+$^{90}$Zr reaction at multiple laboratory angles. Quasi-elastic excitation function and  barrier distributions derived therefrom  have been compared with the Coupled Channels (CC) calculations using CCFULL code. Considering $^{28}$Si as an inert nucleus, and taking into account only vibrational couplings of $^{90}$Zr, the calculations have been found to deviate significantly from the experimental data. In further CC calculations including rotational couplings of $^{28}$Si, the quadrupole deformation was varied in a large parameter space from oblate to prolate ($\beta_{2}$ from -0.5 to +0.5). Similarly, the hexadecapole deformation was varied in a wide range ($\beta_{4}$ from -0.4 to +0.4). A Bayesian analysis was carried out to determine the best choice of ground state $\beta_{2}$ and $\beta_{4}$ values for $^{28}$Si. Following all these analysis, $^{28}$Si has been turned-out to be an oblate shaped nucleus with precise values of $\beta_{2}$=-$0.38 \pm 0.01$ and $\beta_{4}$=+$0.03 \pm 0.01$, respectively. The $\beta_{2}$ value for $^{28}$Si obtained in the present work is observed to be in excellent agreement with the earlier reported values obtained using electromagnetic probes--electron-scattering and Coulomb excitation. A precise value of the hexadecapole deformation for the 
$^{28}$Si ground state, along with a value of $\beta_{2}$, consistent with those from the other probes has been determined for the first time using quasi-elastic scattering. The sign and value of the experimental $\beta_{4}$ show a remarkable agreement with those obtained from Skyrme-Hartree-Fock calculations. Thus, the present results further establish quasi-elastic scattering as a potential route to investigate the ground state structure of exotic nuclei using RIBs where beam intensity is of primary concern. We point out here that among all the probes to determine the ground state deformation of a short lived exotic nucleus, the quasi-elastic scattering and the Coulomb excitations are the ones where the exotic nuclei can be used in the form of a beam bombarding an appropriate stable target. However, the ease of performing ``singles" measurements in quasi-elastic scattering, makes it  somewhat superior  to the Coulomb excitation.

We are thankful to Dr. R. K. Choudhury, Dr. B. K. Nayak, and Dr. V. Jha for discussion at various stages of this work. Authors are also thankful to Mr. Raman Sehgal and Mr. Vaibhav Kumar for helping with the ``ANUPAM" supercomputing facility of BARC. We are grateful to the operating staff of BARC-TIFR Pelletron for the smooth operation of the machine.  UG acknowledges the U. S. National Science Foundation (Grant No. PHY2011890).



\end{document}